\documentclass[aps,prd,twocolumn,floatfix,nofootinbib]{revtex4}   

\usepackage{amsmath}    
\usepackage{graphicx}   

\newcommand{\vectornorm}[1]{\left|\left|#1\right|\right|}

\newcommand{\ket}[1]{|#1\rangle}
\newcommand{\braket}[2]{\langle #1|#2\rangle}
\newcommand{\opbraket}[3]{\langle #1|#2|#3\rangle}
\begin{document}

\title{Quantum mechanics as a consequence of discrete interactions}
\author{Gabriele Carcassi}
 \affiliation{Brookhaven National Laboratory, Upton, NY 11973}
\date{November 25, 2007}

\begin{abstract}
Quantum mechanics is usually presented starting from a series of postulates about the mathematical framework.
In this work we show that those same postulates can be derived by assuming that measurements are discrete interactions: that is, that we measure at specific moments in time (as opposed to a continuous measurement that spans a long time interval) and that the system is in general affected by our measurement.
We believe that this way of presenting quantum mechanics would make it easier to understand by laying out a more cohesive view of the theory and making it resonate more with our physics intuition.
\end{abstract}

\maketitle

\section{Introduction}

It is unfortunate that, after more than half a century that quantum mechanics has become a core part of our scientific understanding, it is still surrounded by a cloud of mystery and perceived as strange and nonintuitive. It is true that it does predict behavior that is odd and counter to our intuition, but does that have to imply we are bound to feel like something escapes us?

Let us compare, for example, with another theory that also has strange consequences, such as the concept of spacetime, time dilation, length contraction or the equivalence of mass and energy. Yet, special relativity is always presented as natural, in fact necessary. We believe that the main difference is that it is presented as coming from simple physical postulates, the principle of relativity and the invariance of the speed of light, which help us make sense of all the other physical consequences.\cite{specrel}

Quantum mechanics, with its uncertainty principle, interference and probabilistic predictions, is usually derived instead from a set of mathematical postulates.\cite{qm} Nothing tells us why a Hilbert space must be used as the phase space, or why observables are associated with operators: that is the starting point. The mathematical results derived from the postulates need to be subsequently interpreted physically, with nothing else to connect them together but the mathematical framework. Should the physics not come first? Should the math not be derived from the physics?

We are left to wonder whether we are missing some sort of "physical postulate" that could help us tie together the theory, that would show us why classical mechanics is insufficient and why quantum mechanics is necessary. We believe that if we were able to present quantum mechanics derived from one or more physical postulates, it would increase our sense of understanding. What is understanding if not being able to identify, in the midst of all that is confusing and misleading, that simple truth from which all others descend?

\section{Looking for a physical postulate}

To reach our goal, our postulate has to satisfy the following four requirements.

First of all, it has to state an obvious physical fact: something that any person who studied quantum mechanics would perceive as plain, even uninteresting. \footnote{After more than half a century that the theory is well established, it is unlikely some important point was missed. And even more unlikely that the author of this paper would be the one to find it!}

Second, it has to imply that classical mechanics is insufficient. It has to tell us why we cannot use it, and, consequently, where we can.

Third, it has to imply quantum mechanics. We need to derive the mathematical postulates that are the starting point of many textbooks:\footnote{For brevity, we omit some more advanced cases, such as the phase space of a composite system or when more than one eigenstate is associated with the same eigenvalues. In addition we are not considering the Schrodinger equation a postulate, in the same way that in special relativity energy-momentum conservation is not a postulate.}
\begin{quotation}
The phase space is a complex vector space where each direction represents a physical state
\end{quotation}
\begin{quotation}
The probability to transition from one (normalized) state to another during a measurement is given by the square modulus of the inner product: $|\braket{\Psi_f}{\Psi_i}|^2$
\end{quotation}
\begin{quotation}
For every measurement there exists a corresponding linear Hermitian operator. The only possible measurement values are the eigenvalues of the operator. The expectation value of the measurement is given by : $\opbraket{\Psi}{A}{\Psi}$. Upon a measurement, the state of the system will transition to the eigenstate associated with the eigenvalue
\end{quotation}

Fourth, and last, it has to imply only quantum mechanics. We should not, for example, derive a theory that reduces to or is equivalent to quantum mechanics. Moreover, since there are currently many interpretations, we need to be compatible with all of them. And since quantum mechanics does not really give answers to such questions as realism or locality, we do not expect to give any either.

In short, there should not be anything revolutionary about either the principle, or any of the items in the derivation.
The novelty should mainly lie in how things are presented, and how they resonate better with our physics intuition.

\section{On describing a system}

At the beginning of the 20th century, experimental and theoretical advances such as the ones by Thompson, Rutherford, Planck and Einstein started to show how both matter and the fields that describe its interactions are made of discrete elements. All physical processes are reduced to interactions among these particles.

Since measurements are physical processes, it follows that they are also interactions. To measure a property of an electron we will need to make it interact with another particle, for example a photon. To measure a property of a photon we will need to make it interact with a charged particle, for example an electron. We have to assume, then, that at the fundamental level \emph{measurements are discrete interactions}. By \emph{interactions} we mean that the system we are studying is going to be affected: more specifically, that some future measurement will change because of our current measurement. By \emph{discrete} we mean that they happen at a specific instant. Two measurements, therefore, will be separated from one another and need to be ordered, which means we will not be able to describe the evolution of the system during a measurement (as it would imply a measurement, an interaction, during another measurement, another interaction): we will only be able to describe the system before and after the measurement. In addition, the fact that measurements are discrete allows us to study them independently of the evolution of the system we are studying, as no evolution will happen during an instantaneous measurement. We can simply turn our attention to the effect of measurement, without studying what happens between measurements and within a measurement.

The fact that \emph{measurements are discrete interactions} will be our physical postulate from which we will derive quantum mechanics. We believe this to be a trivial enough fact to satisfy the first requirement.

Next on our list is to show how this postulate is incompatible with classical mechanics, so let us review some of its aspects. In classical mechanics we define the state of the system we are studying by a set of values that are the result of a few measurements that represent the smallest set from which all the other possible measurements can be calculated. In many cases, for example, position and momentum represent the state, and other measurements, such as energy, can be derived from them. These values will change in time, depending on the external forces that act on the system, creating a trajectory that we usually describe using continuous and differentiable function.

There is an important underlying assumption that allows us to describe a system this way. The fact that we are not required to describe what and when we measure means that we can always conduct a measurement that did not significantly interact with the system. Even when we do describe the effects of a measurement, the fact that we can fully describe the measurement itself means that, at least in principle, we assume there exists another measurement that would not disturb the system and that would allow us to get a full picture. In classical mechanics, then, we are assuming that \emph{measurements are observations}. By \emph{observation} we do not mean what is usually meant in quantum mechanics: we simply mean the intuitive sense of looking at a system without changing it, without interacting.\footnote{Since this is the critical point, we believe the words observer and observable are misleading in the context of quantum mechanics, and that is why we will avoid them. For our purposes, observation is a measurement where the interaction can be neglected, and that does not exist in quantum mechanics.}

In many cases we can disregard the effects of the measurement: if we throw a ball and follow its trajectory, the ball will be interacting with photons whose interactions are not going to significantly change its motion. In this case we can assume that our measurements are observations, therefore we can use classical mechanics to describe the system. But if we are trying to describe the motion of an electron, instead, we cannot simply disregard the effects of the photon interactions: we have to assume that measurements are discrete interactions.

But what exactly changes when we assume that \emph{measurements are discrete interactions}? We explore this question through a thought experiment.

\section{A thought experiment}

Imagine we have a system in an unknown state, and we need to conduct a few measurements to determine its physical state. We assume we conduct one experiment at a time and that nothing is interacting with this system between our measurements. We will not even assume what those quantities are, and assign them the first six letters of the alphabet.

We will start by measuring, A, then B, C, D, E and finally F. If we assume \emph{measurements are observations}, we can go through all measurements, and we determine the full state. In fact, we can even repeat those measurements, just to be sure. We can do this because nothing changes the state of the system, so all those values are guaranteed to be maintained.

\begin{figure}[h]
\begin{center}
\includegraphics[scale=1]{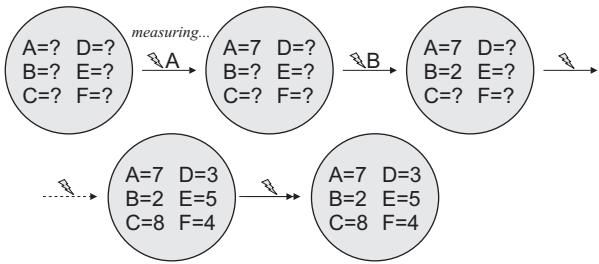}
\caption{\label{fig:thought1}If measurements are observations, we can measure all values one after the other multiple times.}
\end{center}
\end{figure}

If we assume \emph{measurements are discrete interactions}, though, things will be different. We will measure A without a problem, but already with B it could happen that the actual measurement modified the value for A. To be sure we need to measure A again. We can assume that B did not affect A, but, at some point, some measurement will affect some other measurement: the very definition of interaction implies that a future measurement is going to change. For example, let us say we measured A to be 7, and then we proceed to measure D to be 3. When we go back to measure A again, though, we do not see 7 anymore, because the D measurement changed it, and we now measure 6.

\begin{figure}[h]
\begin{center}
\includegraphics[scale=1]{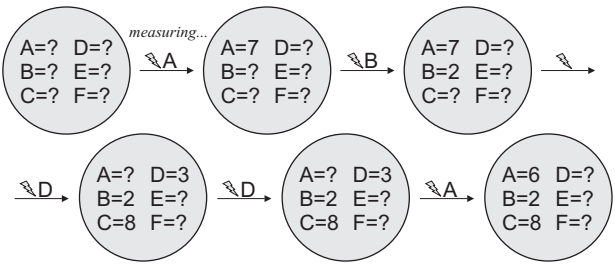}
\caption{\label{fig:thought1}If measurements are interactions, at some point one measurement will invalidate a previous one.}
\end{center}
\end{figure}

Discrete interactions imply that, in general, some measurements are going to affect the outcome of some other measurements. We call these "incompatible". Measuring one essentially invalidates the previous measurement: in our example, when we measured D, we needed to throw away the previous measurement of A because it was no longer valid. This means we need to choose what we want to measure carefully, since measuring one thing means that we will not be able to measure the other. It also means that the order in which we measure different quantities affects the results.

Since incompatible measurements cannot be known at the same time, we can only define the state up to a full set of compatible measurements (where we define \emph{full} as a set in which it is not possible to add a new compatible measurement of which the outcome cannot already be predicted by the measurements already in the set). Those states are the only ones we will be able to distinguish experimentally. Note that this definition also holds for classical mechanics: the only difference is that, in that case, we can include all measurements since they will all be compatible with each other.

Let us now suppose, then, we know the state of a system, and we perform a measurement: what will our predictions be? If the measurement is compatible with measurements that define the state, we will be able to predict the value exactly. But what happens if the measurement is incompatible? The outcome is not part of the state and cannot be calculated from it. Therefore a precise prediction cannot be made. The best that we can hope for is a statistical prediction.

To sum up, discrete interactions imply that there are incompatible measurements, that the state will not contain the results for all possible measurements, but only for a compatible set, and that predictions will be, in general, probabilistic. It should be clear now how discrete interactions are not compatible with classical mechanics, as they force us to redefine our concept of state. And it should also be evident how we have already found many of the core characteristics of quantum mechanics.

\section{On interpretations}

We want to stress that we are not going to make any further assumptions regarding what the state represents and what happens during the interaction.

To be specific, we have not said whether the state also represents some other physical reality as in many interpretations of quantum mechanics. We have not said whether the world is inherently probabilistic and/or God plays dice (as in the Copenhagen interpretation): we only stated that \textbf{we} have to play dice since our prediction will be probabilistic. We have not said whether it is or it is not possible to extend the state with hidden variables that would make the measurement deterministic, even though they themselves could not be measured (as in any hidden variable interpretation, including Bohm-de Broglie\cite{bohm1,bohm2}). We have not said whether many worlds, one for each possible measurements, stem from each interaction (as in the many-world interpretation\cite{many1,many2}). On all of these issues we remain agnostic: they may better describe how and why we have discrete interactions, but we do not actually need anything so specific to proceed.

For us the state, as we already said, just describes all the values that can be measured experimentally (again, this is also true for classical mechanics, with the difference that in that case more can be known). Strictly speaking, though, the state is at least that: anything could be added to make the overall picture more precise in some way. But we stop at this common ground.

\section{On measurements}

Let us see what we can know for sure about measurements.

For each of them, we will start in an initial state and end in a final state. As we said, we will not be able to fully determine the final state from the initial state, so we will have a probability distribution.

Our measurements, though, need to be repeatable.\footnote{One might argue whether repeatability constitutes a separate principle. We note that a non-repeatable measurement would mean a measurement on an isolated system that returns different values every time. Such measurement would be meaningless, because we would not even be able to ascribe it to any particular system. Therefore we believe that measurement itself implies repeatability, or it would not be a measurement.} That is, if we just measured that A equals 7, we need to be able to re-measure 7 with certainty, provided no other interactions changed the state. This means that if we start from one of the possible final states for a measurement, it will not transition to any other state. Or in other words, the probability to transition to itself will be 1, while the probability to transition to any other of the final states will need to be 0.

Another thing that we can say is that in order for two measurements to be compatible, they will need to share the set of final states. This is necessary to be able to have repeatable measurements of both quantities.

One thing that should also be apparent now is that measurements are actually performed on the final state, not on the initial one. When we measure that A equals 7, we cannot really say that A was 7 before we made the measurement. It is only after the transition to the final state that we are sure that A is indeed 7, and that we can repeat the measurement as many times we want and keep obtaining 7. Before the measurement, we do not really know... It could have been 7. Or it could have been 8 and our measurement changed it to 7. It could be that it was not even defined. The point is: we cannot really tell.

This might seem a subtle point, but it is actually changing the picture quite a bit. It means that when we are studying the state evolution, we are, strictly speaking, studying the evolution of our measurements, and not the evolution of the properties of the system we are interacting with. If measurements are observations, the two necessarily coincide; but with discrete interactions they do not. We cannot tell a priori how much of a discrepancy there will be. It is probably going to change from case to case, but in general we should be very careful when interpreting our measurement predictions as describing what the system is doing before we interacted with it.

From this discussion, it seems we will need to be able to keep track of two things. The first is the probability to transition from one state to the other during measurement. The second is the set of all the possible outcomes of each measurement.

\section{The problem with classical probability}

When we ask "what is the probability of transitioning to this particular final state knowing that I start from this particular initial state?" we are posing a question that can be described by conditional probabilities. It is natural to ask whether we can use classical probability to describe them. Let us study a simple case to try to get an answer.

We consider two measurements, A and B, which are to be maximally incompatible. That is, when we measure one, we do not know anything about the second. For simplicity, we assume that both measurements have two outcomes: $a^+$, $a^-$ and $b^+$, $b^-$.

If we consider the event $a^+$, we require that the conditional probabilities are the following: $P(a^+|a^+)=1$ and $P(a^-|a^+)=0$, which is a requirement for the measurement to be reproducible (we always have to measure $a^+$ and never $a^-$); $P(b^+|a^+)=1/2$ and $P(b^-|a^+)=1/2$, which follows from the measurements being maximally compatible (we have an equal chance of getting any B outcome after having measured A). For all four events we will get similar numbers, as shown in Table~\ref{tab:prob1}.

\begin{table}[h]
\begin{center}
\begin{tabular}{|l|l||l|l|}
\hline
\multicolumn{2}{|c||}{$a^+$}&\multicolumn{2}{|c|}{$a^-$}\\
\hline
$P(a^+|a^+)=1$&$P(a^-|a^+)=0$&$P(a^+|a^-)=0$&$P(a^-|a^-)=1$\\
\hline
$P(b^+|a^+)=\frac{1}{2}$&$P(b^-|a^+)=\frac{1}{2}$&$P(b^+|a^-)=\frac{1}{2}$&$P(b^-|a^-)=\frac{1}{2}$\\
\hline\hline
\multicolumn{2}{|c||}{$b^+$}&\multicolumn{2}{|c|}{$b^-$}\\
\hline
$P(a^+|b^+)=\frac{1}{2}$&$P(a^-|b^+)=\frac{1}{2}$&$P(a^+|b^-)=\frac{1}{2}$&$P(a^-|b^-)=\frac{1}{2}$\\
\hline
$P(b^+|b^+)=1$&$P(b^-|b^+)=0$&$P(b^+|b^-)=0$&$P(b^-|b^-)=1$\\
\hline
\end{tabular}
\caption{\label{tab:prob1}Conditional probabilities for two maximally incompatible measurements.}
\end{center}
\end{table}

Given that $b^+$ and $b^-$ describe probability distributions in A, we expect to be able to express them in terms of $a^+$ and $a^-$. We can try to do this by combining $a^+$ and $a^-$ according to classical probability, which means we describe the outcome in which we get either $a^+$ or $a^-$ with equal chance. The conditional probabilities that predict our measurement values given this new outcome will be the averages of the $a^+$ and $a^-$ cases, making all of them 1/2. The problem is that the case where everything is 1/2 is not a state at all: a state is defined by a complete set of compatible measurements, and a state where everything is unknown is not complete, since there is some measurement that can still be taken.

What we actually want to describe when combining $a^+$ and $a^-$ is not, as in classical probability, the case where we get $a^+$ or $a^-$. We want to describe the case where a measurement changed the state so that a+ and a- are now equally likely. We are essentially asking: if after a measurement, A is completely unknown, what have we measured? The answer is: we either measured b+ or b-. We know that by construction. But notice that we have no way of obtaining either of them by combining $a^+$ and $a^-$. If we combine outcomes using classical probability, we can only increase uncertainty, so if we start from A, we cannot get B.

From all of this follows that we need a different kind of probability. One that does not describe uncertainty coming from lack of measurement, but uncertainty coming from incompatible measurements.

\section{Constructing the phase space}

We are now ready to construct the phase space: it will be a vector space in which a vector represents a state of a system. This is similar to classical mechanics, where each point in the phase space represents a state, but this is the only thing the two will have in common.

Since we want to keep track of probabilities, we will use the metric of our space to describe them. To be more specific, we will want the norm of a vector to have to do with the probability of being in the state represented by that vector; we will also want the scalar product between two vectors to have to do with the conditional probability between the two associated states.

First of all, we should understand what orthogonal directions represent, since this will also define the base and dimensionality of our space. As we said, we want the scalar product to be connected to the conditional probability. We also noted that the conditional probability between two outcomes of the same measurements is either 1, if they are the same outcome, or 0, if they are different. This means that we will represent different outcomes of the same measurements by orthogonal directions in our space. If we have a set of compatible measurements, each orthogonal state will represent a definite outcome for all of them, and if the set is complete, the orthogonal states will describe all the possible orthogonal directions. In other words, they will be a base for our space. Any other vector in that space will then be a probability distribution across those states.

We need to be more precise on how exactly we represent the conditional probabilities in our space. In a euclidian space the norm is given by:\footnote{For now, we are indeed considering a real vector space, as nothing has yet told us this is not suitable}
\begin{equation*}
\vectornorm{v}^2 = x_0^2+x_1^2+x_2^2+...
\end{equation*}
That is good for describing lengths, because that is how orthogonal lengths sum. When we sum probabilities of orthogonal outcomes, though, we simply sum the probabilities, so we require that
\begin{equation*}
P(v|v) = P(x_0|v) + P(x_1|v) + P(x_2|v) + ...
\end{equation*}
This means that the probability of being in the state represented by that vector will be the square of the norm. A proper physical state, then, should be represented with a vector of norm 1, to represent the fact that we are certain the system is in that state. Each component, instead, will represent the square root of the probability of measuring the state represented by the base direction:
\begin{equation*}
x_i = \sqrt{P(x_i|v)} = v\times e_{x_i}
\end{equation*}
The component is simply the scalar product with the unit vector along the appropriate axis: the scalar product will represent conditional probabilities.

From the discussion above, we note that any vector in the same direction with a norm less than one is still going to represent the same physical state, just with a different probability of measuring that state. A physical state, then, is really a direction in the phase space, so any unit vector multiplied by any constant still represents the same physical state. As of now, it needs to be between 0 and 1, since it just represents a probability. Later we will make use of this constant to represent quantities that span outside that range.

To recap, the dimension of the phase space will be the number of all possible combinations of the outcomes of a complete set of compatible measurements. A unit vector in that space will represent a normalized probability distribution, and the component along each direction will tell us the square root of the probability for that outcome.

\begin{figure}[h]
\begin{center}
\includegraphics[scale=1]{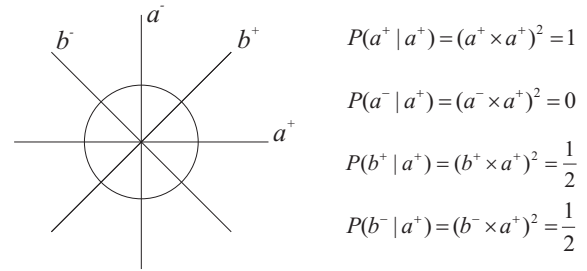}
\caption{\label{fig:phasespace}Phase space for two incompatible measurements.}
\end{center}
\end{figure}

Let us go back to the previous example and have a look at how it all works: $a^+$ and $a^-$ are going to be orthogonal to each other, and the plane they create will describe the probability space for A. What about B? We know that the component for $b^+$ will need to be $\sqrt{1/2}$ on both $a^+$ and $a^-$, which means that $b^+$ is going to be at 45 degrees. $b^-$ will need to be orthogonal to $b^+$, so it will need to be at 135 degrees.

Note that if it were classical probability, we would actually have four orthogonal directions: $a^+b^+$; $a^+b^-$; $a^-b^+$; $a^-b^-$. This would also be the case if A and B were compatible. But the fact that they are incompatible makes them live on the same plane, at 45 degrees from each other.

So, what does it mean to describe the state in this space? It means that at any point in time, the state vector is pointing toward a linear combination of A and B. That is the direction that points toward the measurement that is certain. The orthogonal directions, instead, represent measurements that we know that cannot happen. All that is in between are measurements that are going to be uncertain, with the maximum uncertainty at 45 degrees. This is exactly the kind of uncertainty we need to describe.

\section{On spin}

A careful reader already familiar with quantum mechanics will have probably noticed that the space we constructed in our example is actually the Z-X plane for a spin 1/2 system (we are disregarding Y). $a^+$ and $a^-$ represent $z\uparrow$ and $z\downarrow$, while $b^+$ and $b^-$ represent $x\uparrow$ and $x\downarrow$ states.

Notice one important thing: $z\uparrow$ and $z\downarrow$ are orthogonal in this space while they are 180 degrees apart in the physical space; $z\uparrow$ and $x\uparrow$ are at 45 degrees, while they are at 90 degrees apart in physical space. This should sound familiar since whenever we calculate spin probabilities, half of the physical angle always comes into play. What we are actually using is the angle in the phase space, not the physical angle; they just happen to be related. Another detail: we know that for a spinor we need to make two full turns in physical space to come back to the same state. Again, in the phase space we are actually making just one turn, but at half a turn, we are coming back to the opposite direction, which represents the same physical state. The angle described by the spinor is not an angle in physical space, but is an angle in phase space.

It should be clear by now that we are really heading toward the right direction: we are finding quantum mechanics.

\section{Generalizing our phase space}

We saw how our simple space worked for 2 measurements and 2 outcomes. We should now generalize to the case of $n$ outcomes. Since we will not be able to visualize that through a pictorial representation, we will let the mathematical representation guide us.

In the simple case, we had the vector components expressed in A and B linked by a linear transformation (a 45-degree rotation):
\begin{align}
\begin{bmatrix}
b^+ \\
b^- \hfill
\end{bmatrix}
&=
\begin{bmatrix}
 \sqrt{2}/2 & \sqrt{2}/2 \\
-\sqrt{2}/2 & \sqrt{2}/2 \hfill
\end{bmatrix}
\begin{bmatrix}
a^+ \\
a^- \hfill
\end{bmatrix}
\end{align}

The square of the components represents the actual probability distribution. Note that the matrix needs to be unitary (or one probability distribution would not sum to one) and the elements represent the scalar product between the unit vectors of the different bases. In the case where there are $n$ outcomes, the vector will have $n$ components, and the matrix will be $n\times n$:
\begin{align}
\begin{bmatrix}
b_1 \\
b_2 \\
\vdots \\
b_n \hfill
\end{bmatrix}
&=
\begin{bmatrix}
e_{b_1}\times e_{a_1} & e_{b_1}\times e_{a_2} & \ldots & e_{b_1}\times e_{a_n} \\
e_{b_2}\times e_{a_1} & e_{b_2}\times e_{a_2} & \ldots & e_{b_2}\times e_{a_n} \\
\vdots & \vdots & \ddots & \vdots \\
e_{b_n}\times e_{a_1} & e_{b_n}\times e_{a_2} & \ldots & e_{b_n}\times e_{a_n} \hfill
\end{bmatrix}
\begin{bmatrix}
a_1 \\
a_2 \\
\vdots \\
a_n \hfill
\end{bmatrix}
\end{align}
If we assume that these are still fully incompatible measurements, all the conditional probabilities between A states and B states will need to be $\sqrt{1/n}$, which means we need to have an $n\times n$ matix, where the modulus of all components needs to be the same, but the determinant is non-zero. This is of course not possible if the matrix is made of real numbers.

It is important to understand what this means physically. Suppose a matrix of real numbers represented the relationship between A and B, and we were given a probability distribution of A: we would be able to determine the probability distribution of B. But this is not possible by construction: a uniform distribution for A could be had for any of the different B measurements. We were able to use a real matrix in the simple $2\times 2$ case because we used a minus sign.

We can do something similar if we use complex numbers and use the phase in the same way we used the minus sign: after all, a minus sign is a phase of 180 degrees. This means, though, that we need to extend all of the phase space to complex numbers. First of all, we need to redefine the scalar product by using, instead of the square of the components, the components times their complex conjugates:
\begin{equation*}
\textbf{v}^*\textbf{v} = x_0^*x_0+x_1^*x_1+x_2^*x_2+... = \braket{v}{v}
\end{equation*}
Also, the probability distribution is a vector of complex numbers. But this should not throw us off: it is still a probability distribution. In fact, it is two probability distribution in one. To convince ourselves of this fact, let us look again at that system of equations: it is n complex linear equations, that is $2n$ real linear equations. If we fix both probability distributions for A and B, we are fixing the modulus for both distributions, while leaving the phases undetermined. This leaves us with a system of $2n$ equations in $2n$ real unknowns, which means that we will be able to determine the phases from the probability distributions. What happens is that, while the modulus is the probability distribution of one measurement, the phases describe the probability distributions of the incompatible measurements. What exactly each phase is, we cannot say in general: it depends on the actual matrix transformation. But the overall point remains valid: the vector state is actually two probabilities combined.

This is an extremely useful property to have. Physically it will never make sense to operate on a probability distribution without operating on the distributions of all incompatible measurements. This is why we constructed the space this way. By having the two distributions so combined, when we write an equation involving one we are actually writing an equation involving the other, too. The careful reader will have recognized that in the basis of the position measurement, this is nothing other than the wave function. The modulus is the probability of measuring the system at a given position, while the phase will give us information about the maximally incompatible measurement, its momentum. While making the relationship more precise is outside of the scope of this work, we want to underline how "wave function" is actually a poor name, as it too close to the wave interpretation. If we could rename it, it would be something along the lines of ``multi-probability distribution''.

In any case, we have finally arrived at the first mathematical postulate: the phase space is a complex vector space. If we need to represent a continuous measurement, such as position or momentum, we will make the number of measurements go to infinity, and the distance between them go to zero, in which case the components of our state will be represented by a complex function.\footnote{Note that, up to now, we have avoided the use of Dirac notation: this was deliberately done. We believe that, by using the familiar vector space notation as much as possible, the quantum phase space is introduced in a more familiar way.} We have also have deduced the second postulate: the probability to transition for two normalized states is given by the square modulus of the inner product
\begin{equation}
|\braket{\Psi_a}{\Psi_b}|^2
\end{equation}

\section{Measurement representation}

The last thing we need to do is define how to describe the measurement itself. Ideally, we want to be able to keep track of all possible final states and of all the possible values of the measurement. Since our description is statistical, we will also need some way to work with expectation values.

There is a very convenient way to  describe all of these by using the fact that we can multiply a state vector by a constant, and it still represent the same physical state. The overall idea is to multiply states by the measurement value, so that the metric of the vector is not only the probability, but the probability times the measurement value, which will give us the expectation value. For every measurement, then, we will have a corresponding operator that changes the vector in phase space. But we have to be careful to avoid misconceptions: this operation does not represent the state change that happens during measurement. What it does represent is the transformation from the vector that represent the probability distribution to a vector that represents the expectation values. Let us see how this works in detail.

\begin{figure}[h]
\begin{center}
\includegraphics[scale=1]{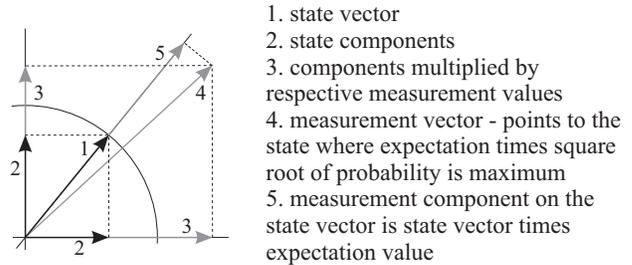}
\caption{\label{fig:measurerep}The measurement operator multiplies each state by the expectation value of the measurement.}
\end{center}
\end{figure}

We can start from the simplest case, which is a vector that represents one of the final states of the measurements. In this case, we will simply multiply by the associated measurement value. Since the norm of the state vector is one, the product of the measurement vector and the state vector will be the measurement value.
\begin{subequations}
\begin{align}
A\ket{\Psi_i}&=a_i\ket{\Psi_i}  \\
\opbraket{\Psi_i}{a_i}{\Psi_i}&=a_i\braket{\Psi_i}{\Psi_i}=a_i
\end{align}
\end{subequations}
Mathematically, this will mean that the final states of a measurement are represented by the eigenvectors of the operator: an eigenvector is a vector that the operator changes only in the norm, and not in the direction. The measurement value, then, will be the eigenvalue associated with that eigenvector.

Since the expectation is a linear operator, we will require that the operators associated with our measurements will also be linear. This means that for a vector that is not a final state, each component along the directions of the final states will be multiplied by the eigenvalue associated with that state. Since the value will be different along each direction, a state that is not a final state will point in a different direction. So, only the final states associated with the measurement will be eigenstates.\footnote{To be absolutely clear: the fact that the final states and only the final states are eigenstates of the measurement operator does not represent the fact that the state is not changed by the measurement. The two statements are both true, but they are unrelated. The first statement is represented mathematically by the fact that the final states are orthogonal, while the second represents the physical fact that the final states are associated to a well defined measurement and not to an expectation value.}

What is even more interesting is what happens to the norm of the vector. We know that the state vector is the sum of all outcomes weighted according to their probabilities\footnote{The relationship is slightly looser: to be precise, it is actually a complex number that multiplied by its complex conjugate returns the probability.}
\begin{equation}
\ket{\Psi} = c_0\ket{\Psi_0} + c_1\ket{\Psi_1} + ...
\end{equation}
When we apply our measurement operator what we have are all the possible measurement values weighted according to their probability.
\begin{equation}
A\ket{\Psi} = a_0c_0\ket{\Psi_0} + a_1c_1\ket{\Psi_1} + ...
\end{equation}
If we take the inner product with the original state vector, we will end up with the expectation value
\begin{equation}
\opbraket{\Psi}{A}{\Psi} = a_0P_0 + a_1P_1 + ...
\end{equation}

This is indeed a powerful way to represent measurements. There are other nice properties the operators share, which come into play when using them in equations, but we will not digress since our point here was simply to reach the third mathematical postulate, which we did.

\section{Conclusion}

We have seen how we can derive quantum mechanics from the simple physical postulate that all \emph{measurements are discrete interactions}: this is the main difference introduced by this theory. The fact that there are incompatible measurements, that our state is only defined up to a set of compatible ones, that our predictions will be probabilistic, that we need a complex vector space as our phase space and that we associate operators with each measurement are all consequences. All the rest that can be derived from these, such as the uncertainty principle or interference, will be consequences as well.

We believe that presenting quantum mechanics in this way helps develop a more intuitive physical understanding and makes it feel less arbitrary and more necessary. We saw that the probabilistic nature of our predictions are basically linked to our inability to measure, and therefore distinguish between, states that differ from each other only by incompatible measurements. We saw that quantum probability is very different from classical probability, because while the latter deals with uncertainty coming from lack of measurement, the former deals with uncertainty coming from impossibility of measurement. The discussion on the two-direction spin system is very instructive because it allows us to visualize the phase space very clearly, which makes it easier then to generalize to more complex systems. And finally we saw how the operators associated with measurements are actually not describing a transformation of the state, as a rotation or a translation would, but they are merely multiplying the state by the measurement values. These are all small but important points that help us sharpen what physics we are describing and how it is represented in our math.

We do not think this is the whole story, though. As $c$ is a very critical constant in special relativity, $\hbar$ is in quantum mechanics and we would expect it mentioned in the physical postulates. This is probably a hint that we are still missing a piece.

Nonetheless, we believe that this derivation is a good step forward in our understanding of quantum mechanics and a useful tool when introducing the subject.

\end{document}